\documentstyle[aps,eqsecnum,preprint]{revtex}
\begin{document}

\preprint{MSUCL-973}
\title{Causality Violations in Cascade Models of Nuclear Collisions}

\author{G. Kortemeyer, W. Bauer, K. Haglin, J. Murray and S. Pratt}
\address{
NSCL/Cyclotron Laboratory, Michigan State University, East Lansing, MI
48824-1321, U.S.A.}
\date{\today}
\maketitle

\begin{abstract}

Transport models have successfully described many aspects of intermediate
energy heavy-ion collision dynamics.  As the energies increase in these
models to the ultrarelativistic regime, Lorentz covariance and causality
are not strictly respected. The standard argument is that such effects are not
important to final results; but they have not been seriously considered at
high energies. We point out how and why these happen, how serious of a problem
they may be and suggest ways of reducing or eliminating the undesirable
effects.

\end{abstract}

\narrowtext

\section{Introduction}
Experiments like the upcoming nucleus-nucleus
collisions at Brookhaven's Relativistic Heavy Ion Collider will without doubt
open new possibilities to study the properties of nuclear matter at extreme
pressures and temperatures. Of special interest is the expected phase
transition between hadronic and quark-gluon matter. The interpretation of
these complex collisions and the possible generation of a quark-gluon plasma
poses a major problem: what are the experimental signatures? In an effort to
aid the answering of this question a theoretical model for the collision
processes that goes beyond a phenomenological description must be developed.
One possible microscopic approach is to extend the semiclassical transport
theory to high-energy
physics\cite{mosel,greiner1,greiner2,geiger1,geiger2,geiger3,geiger4,geiger5}.

Simulations of ultrarelativistic heavy ion collisions inevitably
involve both soft and hard processes.  Below some energy scale, soft or
nonperturbative phenomena necessitate phenomenological description.
Uncertainty is duly noted but for the foreseeable future a rigorous theory
for soft physics is beyond understanding.  However, throughout much of the
collision hard processes dominate which are described by elementary
interactions between quarks, antiquarks and gluons (partons) as essentially
semiclassical particles.  After parton initialization according to some
reasonably chosen nucleon structure functions, spacetime propagation
is accomplished by discretizing time into units $\Delta
t_{\mbox{\scriptsize step}}$ and updating phase space densities according to
relativistic transport equations including a crucial collision term.

Scattering processes in this theoretical framework are assumed to be
non-retarded which, on a microscopic level, leads to information transport
with velocities that can approach $\sqrt{\frac\sigma\pi}/\Delta
t_{\mbox{\scriptsize step}}$, where $\sigma$ is the parton-parton cross
section. That
is, it can increase without reasonable bound. As energies for the individual
interactions decrease the corresponding cross sections increase. The timestep
$\Delta t_{\mbox{\scriptsize step}}$ is for reasons of convergence chosen to
be less than the parton mean free paths divided by their velocities: of the
order a few thousandths of a fm/$c$. Two problems occur immediately: On a
macroscopic level a series of subsequent causality violating interactions can
lead to shock-waves propagating faster than the speed of light. This is clearly
unphysical. On a microscopic level the time-ordering of the incoming and
outgoing partons of a scattering process becomes frame-dependent and Lorentz
covariance is lost -- this also is unphysical. These two problems are
especially serious in ultrarelativistic parton cascades since collision rates
are high, and the mean free paths approach the interparticle distance. Problems
arise from describing quantum-dynamical processes in a semi-classical picture,
and from the demand of Lorentz-invariance in an equal-time-character
simulation. Where is an acceptable compromise? And what does acceptable mean?

These are the rather technical questions we address in this study, while in
\cite{bauer2} we are more focussing on the physics of parton cascade codes.
Our paper is organized
in the following way: In Sect.~\ref{sec2} we describe in some detail
the origin of superluminous information transport on a macroscopic scale
and suggest some first steps to eliminate it. Then in Sect.~\ref{sec3}
we move toward microscopic physics and provide mathematical details
for origins of unphysical effects.  Whenever cross sections are finite and
action-at-a-distance influences particle trajectories these problems
inevitably arise. Sect.~\ref{sec4} deals with the resulting frame-of-reference
dependence of the simulation. In Sect.~\ref{sec5} we describe our version of a
parton cascade implementation.  At this stage we have only initialization and
temporal development through much of the hard physics. Later stages including
hadronization are ignored since they are outside the scope of this work. Then
in Sect.~\ref{sec6} we compare several different schemes which reduce or
eliminate superluminous transport. Methods include simply blocking collisions
or truncations, scaling the cross sections downward while increasing the
number of particles, and suppressing low-energy collisions. We also consider
so-called wee-partons as a different way to define the initial conditions of
the simulation, and point out how much this scheme affects causality violating
mechanisms. Finally, in Sect.\ref{sec7} we conclude by briefly summarizing, and
by discussing the outlook for future studies.

\section{Macroscopic Causality Violations}
\label{sec2}
Superluminous macroscopic information transport occurs
mainly in the transverse (perpendicular to the beam) direction. The signal
can travel over the diameter of the cross section
$\sigma$ in a single timestep and then continue this propagation from
timestep to timestep. Transverse signal velocities can therefore reach
$\sqrt{\frac\sigma\pi}/\Delta t_{\mbox{\scriptsize step}}$ over several
timesteps, one is reminded of a chain of falling dominoes. The situation is
depicted in figure \ref{fig1} for such a transport.

This is a general problem of all transport
codes, and is worsened because the gluon-gluon cross section becomes rather
large for low energies (instead of vanishing). Along with the large gluonic
cross sections are relatively large gluon densities which result in very
high probabilities for scattering in subsequent timesteps. Therefore
information transport could be supported over relatively large distances
without damping.

In many existing cascade codes first steps for dealing with the problem of
superluminous signals have been taken: Most of the codes, including ours, allow
only one interaction per particle per timestep. This restriction is consequent
and clearly justified
since a timestep is the shortest scale in the model. This
restriction prevents signals from avalanching over huge volumes within only
one timestep.

A second restriction implemented in many codes is the ``closest approach''
criterion. For a scattering process it demands in addition to ``spatial
distance within total cross section'' also ``the two particles have reached
their point of closest approach assuming their current trajectories''. Usually
one would look at $|x^\mu x_\mu|$ with $x_\mu$ being the four-vector distance
between the two particles involved, and demand that this quantity is minimal.
As the two particles' positions are taken in the same timestep this quantity
reduces to the spatial distance squared. Even though this looks like a
Lorentz-invariant criterion, it is not, since the conceptual necessity of
taking both particles' position at the beginning of a timestep (i.e. with
vanishing time-separation in the lab-frame) is a non-Lorentz invariant
restriction, see section \ref{sec4}.
Nevertheless, the ``closest approach'' restriction prevents causality violating
signal transport in the longitudinal (parallel to the beam) direction, but
unfortunately has no effect on the transverse direction.

To demonstrate those causality violating shock-waves, the parton cascade
simulation of a 100 GeV/nucleon (p,Au)-collision was run with different
constant cross sections and $S$-wave scattering. Fig.~\ref{fig2} shows the
distance of scattering events from the
beam axis versus the simulation time for different
cross sections. While during the initial stages of the collisions the outmost
scattering events occur at distances larger than those allowed by causality
arguments, the information transport soon appears to be damped. It turns out
that this
phenomenon is not -- as one would expect -- mainly due to a dropping of the
collision rate, but rather to a form of random-walk: only a fraction of the
individual signal propagations lead outwards, others have the opposite effect,
which results in an effective damping. In fact it turns out that the
expectation value of the distance of scattering events from the beam axis is
roughly proportional to the square-root of the simulation time, which is a
characteristic feature of random-walk mechanisms, compare the top panels of
Fig.~\ref{fig2}. This mechanism is a valid description until the cross sections
are so large that the parton distribution basically appears solid, in that
limit the information travels outwards proportional to the simulation time, see
the panel for $\sigma=0.5$ fm$^2$. By the nature of random-walks, the
information expands
fastest in the initial phase of the collision. With realistic energy-dependent
cross sections however, those initial scattering events happen at lower cross
sections than the ones in the later stages due to the higher c.m.-energy,
which partly suppresses this initial outburst. Overall for $\sigma$=0.2 fm$^2$
the causality violations are still rather moderate, for 0.5 fm$^2$ the effect
becomes dominant.

Fig.~\ref{fig2} also shows the maximum distance of scattering events from the
beam axis as they occurred for different cross sections. As it turns out, this
value depends linearly on the cross section for an extended range, beyond
$\sigma\approx0.4$ fm$^2$ the
outmost scattering events occur at a distance that can only be reached with
superluminous signal velocities.
\section{Microscopic Causality Violations}
\label{sec3}
In addition to macroscopic causality violations such as superluminous
shock-waves, both the non-retarded interactions and the model inherent
propagation of
the whole particle configuration from one timestep to the next lead to
causality violations on the level of elementary scattering processes.

A Lorentz-invariant simulation of parton scattering would require a truly
4-dimensional configuration space, in which an interaction can be established
between any space-time point of one parton's trajectory and any other
space-time point on another parton's trajectory. In the framework of classical
fields, the scattering of two particles would be represented by their
continuous
change of trajectory in the retarded field of the respective other particle.

The actual realization of scattering events in the simulation only agrees in a
certain limit with this model, namely if the two partons are coming from
infinity with infinite rapidity and opposite directions:
with increasing rapidity, in the lab-frame the fields are distorted to
pancakes with an orientation perpendicular to the trajectory, reducing the
dominant part of the interaction between the particles to a shorter and shorter
region along their trajectories. In the limit of infinite rapidity, the
interaction between two partons travelling with a non-vanishing
impact parameter in opposite directions would be reduced to a momentary
interaction at the point of their closest approach, making them change their
trajectories at equal-times. This interaction would take place at a spacelike
distance equal to the impact parameter -- why does this not violate causality?
The reason is, that the parton's field-pancake -- even though it is travelling
along with its present position -- is build up from contributions
out of the parton's past, the respective other parton does not scatter from the
field generated at the first parton's present position, but from a field
component that was generated along the first parton's trajectory at a lightlike
distance.

This limit agrees both with the claim of Refs.~\cite{greiner1,greiner2} that
the interaction distance should be spacelike, because otherwise an interaction
would influence the absolute past of one of the partons, and with the actual
realization of scattering processes in the simulation: in the framework of a
3+1 dimensional transport simulation the distance of the testparticles is by
the model itself determined to be spacelike: The information available at the
beginning of a timestep is no more than the points of the trajectories on a
certain $x,y,z$-hyperplane, as well as the corresponding momentum 4-vectors of
the particles, also only at that time-coordinate. The lightcone of any one
particle extends both before and after that hyperplane, but has no extension
into that hyperplane itself, causing all other particles present in the
simulation to have a spacelike distance from it.

But this limit soon loses its applicability: Staying in the model of
interactions between the partons due to classical fields, as already pointed
out, the field travelling with the partons is in fact build up from
contributions out of their history, which for partons coming from infinity
results in Lorentz-contracted field-pancakes. But in a cascade
simulation it is highly  unlikely that seen from a parton $A$, parton $B$
already was on the same trajectory a lightlike distance ago. In fact, as the
partons are nearly travelling with the speed of light themselves, spacetime
points with lightlike distances along their trajectories can happen to be very
long ago, which in connection with the high interaction rates makes the
idealized picture of the field pancakes inapplicable. Changes in trajectories
can make more than one point along the trajectory of parton $B$ have a
lightlike distance to parton $A$, the acceleration connected to the changes
in trajectory of parton $B$ generates additional fields. Even worse, through
the mechanism of parton generation and absorption, parton $B$ might not even
have existed at lightlike distances from parton $A$, still within the
framework of the transport simulation, scattering would be possible -- and
causality violating.

Also, as in the idealized picture the c.m.-frame is moving parallel to the
partons' trajectories and perpendicular to their distance at the point of
closest approach, the equal-time character is true in both the lab- and the
c.m.-frame. As soon as the two partons are not travelling into opposite
directions, their spacial distance also leads to a time-separation, making the
scattering time different in the lab- and c.m.-frame.
\section{Frame-Dependencies and Time-Ordering}\label{sec4}
An obvious consequence of the model's causality violations is its
frame-of-reference dependence: The simulation will for example certainly
lead to different results when run in the rest-frame of the target.
This was already shown for the example of a ${}^{12}C+{}^{12}C$ internuclear
cascade
calculation in Ref.~\cite{kodama}, the authors demonstrated that both the total
number of collisions and
the individual time-ordering of given collisions were frame-dependent.
As obvious that seems from the simple problems pointed out in sections
\ref{sec2} and \ref{sec3}, for a parton cascade it is not yet the whole story:
The simulation cannot even be run in the rest-frame of any one nucleus,
because there is no possibility to create appropriate initial
conditions. The parton distribution in nucleonic matter is frame-dependent, the
higher the energy of a nucleon the higher the number of virtual partons. To
put it in other words, usually components of the nucleonic wavefunction are
considered a parton if they carry a momentum fraction beyond a threshold
given by a minimum $x$-value, the latter one being a parameter of the model.
However, by boosting the nucleonic wavefunction into an accelerated frame of
reference, a procedure described by the model-parameter $Q^2$ of the parton
distribution function $f(x,Q^2)$, more and more components of it cross that
threshold $x$-value, leading to higher and higher parton numbers.

Therefore, when asking questions about how the collision of two partons looks
in their rest-frame as compared to the lab-frame, another valid question is: Do
those two partons even exist in the other frame? Or are in that frame other
partons around that the collision partners would be much more likely to scatter
with instead?

However, as this question cannot be addressed within the framework of a
transport simulation, it is worthwhile to study the effects of a transformation
of two given partons into their c.m.-frame. It turns out that in addition to
the above inconsistencies, by boosting the partons with {\boldmath$\beta$
\unboldmath} into their c.m.-system, there they have a time-separation of
\begin{equation}
\Delta t_{\mbox{\scriptsize cm}}
=-\gamma\mbox{\boldmath$\beta$\unboldmath$\cdot\Delta$\boldmath$r$\unboldmath}
_{\mbox{\scriptsize lab}}\ .
\end{equation}
The partons' positions, being taken at equal time in the lab-frame, will,
unless {\boldmath$\beta$\unboldmath} and
{$\Delta$\boldmath$r$\unboldmath$_{\mbox{\scriptsize lab}}$} are perpendicular,
not be at equal time in their c.m.-frame and vice versa. The construction of a
``Lorentz-invariant CMS distance'' as in Refs.~\cite{greiner1,greiner2} seems
doubtful. This time-separation can very well be larger than the transformed
timestep length, leading to a situation where the partons' positions are
not taken within the same timestep within the c.m.-frame anymore.
As a result, the closest approach criterion can only provide a means to get an
averaged point in time for an actual scattering event.

For this criterion,
described in section \ref{sec2}, one can choose to define it either in the
particles' c.m.-frame or in the lab-frame (a third approach will be
described in subsection \ref{sec6f}).
Because of the non-vanishing
time-separation of the partons in their c.m.-frame, the two possibilities are
not equivalent. We compare both methods and find nevertheless no significant
change in the collision rate, even though the individual collisions happening
in both simulations are different. For our further
simulations we have chosen the c.m.-frame, since intuitively this frame seems
to be more significant for the individual scattering events. In this frame,
the scalar products of the momentum and
the distance vector both at the beginning and at the end of the timestep are
to be calculated. If this scalar product changes sign, the position of closest
approach is reached within this timestep. In figure \ref{fig3}
$\Delta\mbox{\boldmath$r$\unboldmath}_{\mbox{\scriptsize cm}}$ denotes the
distance vector at the beginning of the timestep,
$\Delta\mbox{\boldmath$r$\unboldmath}_{\mbox{\scriptsize cm}}'$
at the end. The criterion is $(\mbox{\boldmath$p$\unboldmath}
_{\mbox{\scriptsize cm}}\cdot\Delta\mbox{\boldmath$r$\unboldmath}
_{\mbox{\scriptsize cm}})(\mbox{\boldmath$p$\unboldmath}
_{\mbox{\scriptsize cm}}\cdot\Delta\mbox{\boldmath$r$\unboldmath}
_{\mbox{\scriptsize cm}}')\le0$.

The non-vanishing c.m.-time-separation unfortunately also makes the
calculation of $\Delta\mbox{\boldmath$r$\unboldmath}_{\mbox{\scriptsize cm}}'$
ambiguous. In our model, the c.m.-positions at the end of the timestep are
calculated by propagating the partons within the lab-frame and boosting the
result into the c.m.-frame. The result, expressed in c.m.-quantities, is
\begin{eqnarray}
\Delta\mbox{\boldmath$r$\unboldmath}_{\mbox{\scriptsize cm}}'
&&=\Delta\mbox{\boldmath$r$\unboldmath}_{\mbox{\scriptsize cm}}\\
\nonumber+&&\left(\frac{\mbox{\boldmath$p$\unboldmath}_{\mbox{\scriptsize cm}}}
{\gamma(E_{\mbox{\scriptsize cm,1}}
+\mbox{\boldmath$\beta\cdot p$\unboldmath}_{\mbox{\scriptsize cm}})}+
\frac{\mbox{\boldmath$p$\unboldmath}_{\mbox{\scriptsize cm}}}{\gamma
(E_{\mbox{\scriptsize cm,2}}-\mbox{\boldmath$\beta\cdot p$\unboldmath}_
{\mbox{\scriptsize cm}})}\right)
\Delta t_{\mbox{\scriptsize step}}\ .
\end{eqnarray}
Due to the spacelike distance of the partons however, the definition of
``beginning'' and ``end of a timestep'' becomes frame-dependent.

Consider a process where the partons $A$ and $B$ scatter and produce a
parton pair $C$ and $D$, see figure \ref{fig4}. In the lab-frame this happens
within one timestep, at the beginning of the lab-frame's timestep partons $A$
and $B$ are present, at the end partons $C$ and $D$. In the transformed
timestep
within the c.m.-frame however, this set-up can easily be distorted to a
situation where in the beginning of that timestep partons $B$ and $C$ are
present, and at the end partons $C$ and $D$. Seen from yet another frame, in
the beginning $A$ and $D$ could exist. One consequence is that in our
simulation re-scattering processes have to be explicitly forbidden:
In spite of the closest-approach criterion re-scattering would still be
possible if two partons are scattered towards each other -- as the distance
between partons involved in a scattering event can very well be greater than
$\Delta t_{\mbox{\scriptsize step}}c$, it is possible for those two scattered
partons to again reach a point of closest approach in a later timestep. In the
case of a retarded interaction, this would not be possible.

More relevant then the time-scale given by the timestep, which is a
model-dependent parameter, is the time-scale given for example by the time it
takes for the two nuclei to cross through each other, which is about
0.2 fm/$c$. This time-interval corresponds to an impact parameter of 0.2
fm beyond which the time-ordering of scattering events is frame-dependent.
As the interactions in question happen between particles travelling into the
same direction, those would also be low-energy interactions with cross
sections very well in the range of such impact parameters. The effects of
this frame-dependent time-ordering of the incoming and outgoing particles of
a scattering event have to be subject of extensive examination.

An interesting approach is that of Refs.~\cite{lang,danielewicz}:
In this method,
configuration space is divided into boxes. Within the boxes the scattering
partners are randomly chosen according to a probability that is a function of
the cross section -- convergence of this method towards
the solution of the Boltzmann equation for an infinitely small box
size and timestep length, as well as an infinite number of test-particles,
is shown in Ref.~\cite{badovsky}. The code of Rev.~\cite{danielewicz} was
successfully made more efficient by only considering a fraction of the possible
pair combinations within each box, and compensating the otherwise resulting
loss in collision rate by an enhanced probability for the collision of the
chosen pairs.
The authors of Ref.~\cite{lang} claim that this stochastic method of
formulating the collision term is covariant since it is
dealing with transition rates instead of geometrical interpretations,
therefore no problems
connected with the time-ordering of processes would occur. In fact, since
in the model the time-order of processes is chosen randomly anyway, the model
has no ``right'' time-order that could be distorted by relativistic effects.
However, this new approach does not overcome the problem of superluminous
shock-waves. Within a given cell of longitudinal size $\delta r_{\parallel}$
and perpendicular size $\delta r_{\perp}$ any two particles have a chance of
colliding. Since the subsequent advection step may carry some of these into
neighboring cells, the maximum transverse velocity for information transport
will be $\delta r_{\perp}/\Delta t_{\mbox{\scriptsize step}}$. Since in general
$\delta r_{\perp}^2$ is several times larger than the cross section $\sigma$,
the resulting maximum possible causality violations in transverse direction are
even larger than in the method discussed above. In addition, the stochastic
method also allows for superluminous information propagation in longitudinal
direction, with velocity $\delta r_{\parallel}/\Delta t_{\mbox{\scriptsize
step}}$.
\section{The Parton Model}\label{sec5}
Classical simulations, which have been successfully applied to heavy ion
collisions at intermediate energies \cite{bertsch}, are now
\cite{mosel,greiner1,greiner2,geiger1,geiger2,geiger3,geiger4,geiger5} being
extended to high energies. The main step in the extension of this microscopic
model is using a parton based picture of the nuclei rather than a hadronic
picture; consequently the interactions between the testparticles are to be
described in the framework of QCD, leading to so-called parton cascades.

Our code works in 3+1 dimensions using fully relativistic kinematics for the
partons, where the quarks are consequently treated as massive particles. Both
quarks and gluons can be off-shell. The initial conditions are
determined by standard parton distribution functions $f(x,Q^2)$ \cite{cteq},
where the value for $Q^2$ and the minimum $x$ are parameters of the model.
Technically stability of the incoming nuclei is guaranteed both by a coherent
motion of all partons in longitudinal direction, and by the restriction
that particles from the same nucleus cannot scatter with each other before at
least one of them has scattered with a particle from the other nucleus.

In this preliminary version of the code only QCD processes with two partons in
the incoming and two partons in the outgoing state are implemented
\cite{bauer2,bauer1}. Phenomenological screening or cut-off masses have been
added into the propagators to avoid divergent total cross sections. However,
the gluon-gluon scattering cross section includes a four point diagram which
does not contain a propagator. This means that the
divergence in this cross section must be handled differently from the other
cross sections. One usually regularizes the gluon-gluon cross section by
setting it to a constant value below a certain cut-off energy, in our case we
have chosen a cutoff of $s\approx0.25$ GeV$^2$ corresponding to
$\sigma\le0.45$ fm$^2$. This cutoff seems appropriate because it agrees both
with a reasonable c.m.-energy below which Perturbative QCD loses
validity, and with the limiting cross section for superluminous shock-waves
found in section \ref{sec2}. Other methods of cutting off these cross sections
while maintaining the correct physics are being investigated. Also, no medium
modifications to the elementary cross sections are taken into account.
Fig.~\ref{fig5} shows the cross sections being used. Our analytic expressions
for cross sections have been checked in the massless quark limit against
results from the literature \cite{combridge}, some of them also in the massive
case \cite{eichten}.

\section{Comparison of Methods for Dealing with Superluminous Signals}
\label{sec6}
This section contains a comparison of different methods for dealing with the
problem of superluminous signal transport and is the outcome of simulating
central collisions of a 100 GeV proton on a 100 GeV per nucleon gold nucleus.
This set-up seems suitable since it allows observation of the information
transport from
the incoming (comparably small) particle in nuclear matter. Table~\ref{tab1}
summarizes the parameters used for the simulations. The timestep length of
0.0002 fm/$c$ was determined by running the simulation with different timestep
lengths and observing the total number of collisions. It turned out that the
number of collisions did not change significantly below the value chosen
anymore, even though the simulation does not necessarily get
more precise with smaller timesteps due to the causality violating effects.
With the values chosen for $Q^2$ and the minimum $x$-value approximately 9000
partons per testrun are generated. For the following
calculations, the longitudinal extent of the nuclei in our model is determined
by the Lorentz-contraction only, Fig.~\ref{fig7} shows the initial
configuration in the lab-frame and 1 fm/$c$ later. The
effects of ``wee-partons'' are examined in the last subsection. The cross
sections were energy-dependent according to section \ref{sec5} and
Fig.~\ref{fig5}.

The simulation was first run for $S$-wave scattering in the expectation that
this would be the worst case. Compared to the realistic forward peaked angular
distributions, $S$-wave scattering favors transverse signal propagation.
However, it turned out, that the angular distribution hardly makes any
difference as far as causality violating effects are concerned, the only
noticeable change was in the rapidity distributions, where it turned out
that the gap in rapidity between unscattered and scattered partons was wider
for $S$-wave scattering: the partons lose more units of rapidity in the
initial collisions. The small effect of the angular distributions is
understandable from the fact that even isotropic distributions in the
c.m.-frame are strongly forward peaked in the lab-frame.

The top row of Fig.~\ref{fig8} summarizes the results for a simulation with a
realistic angular distribution. The top left histogram shows the signal
velocity
distribution regarding directly subsequent collisions, see Fig.~\ref{fig1}.
In other words, with $r^\mu_{n,i}$ being the position of the $i$th collision of
the $n$th particle, the distribution of
\begin{equation}
v_S(n,i):=|\mbox{\boldmath
$r$\unboldmath}_{n,i}-\mbox{\boldmath
$r$\unboldmath}_{n,i-1}|/(t_{n,i}-t_{n,i-1})
\end{equation}
for all $n$ and $i>1$ is shown.
The solid histogram shows the radial component of those velocities, the dashed
histogram includes the longitudinal component -- the strong peak at $c$ results
from processes for which the signal transport was dominated by the partons'
motion with the respective nuclei. On the order of 50 isolated scattering
events with velocities from 50$c$ up to 1360$c$ have been suppressed in the
plot.

The second panel shows a more general impression of the signal velocities.
Instead of just taking into account the signal velocity occurring within
one single scattering event, in this histogram the signal velocity from the
very first to the very last scattering event that a parton was involved in
is calculated. With the notation from above and $N(n)$ being the number of
collisions that the parton $n$ was involved in, the distribution of
\begin{equation}
v_A(n):=|\mbox{\boldmath $r$\unboldmath}_{n,N(n)}-\mbox{\boldmath
$r$\unboldmath}_{n,1}|/(t_{n,N(n)}-t_{n,1})
\end{equation}
for all $n$ is shown.
Again, the solid histogram shows the radial component of the
signal velocities, the dashed histogram also takes into account the
longitudinal components.

Because of damping effects the average velocities are much smaller than the
velocities from one event to the next, the third panel shows how
subsequent scattering events are leading to this overall damping of peak
velocities occurring in single scattering events: again starting from the
very first scattering event of a parton, the signal velocity to the
$i$th following scattering event of the partons
\begin{equation}
v(n,i)=|\mbox{\boldmath
$r$\unboldmath}_{n,i}-\mbox{\boldmath
$r$\unboldmath}_{n,1}|/(t_{n,i}-t_{n,1})
\end{equation}
is calculated, and the
average $\langle v\rangle(i)$ over all particles $n$ that had an $i$th
collision is shown; again both
the total and only the radial component of the velocities are plotted.

Finally, the rightmost panel shows how information is travelling outwards.
The outer
boundary of its spreading is after a very fast expansion in the first few
tenth of a fm/$c$ travelling with approximately 0.17$c$. The maximum distance
of a scattering event from the beam axis is 2.5 fm. A more detailed analysis
of the overlay of the average c.m.-energy and the distance of scattering events
from the beam axis shows that the expansion of a shock-wave begins rapidly
immediately after the c.m.-energies get smaller, but then is damped out -- only
the initial stages of the collision seem to be problematic. However, one should
note that in the later stages the individual collisions happen with very high
signal velocities, and that it is rather by the random-walk character of the
signal propagation over longer distances that the information does not travel
with the same high velocity, compare Sec.~\ref{sec2}.

\subsection{Restrictions on the Signal Velocity}
An obvious way to avoid superluminous signal velocities is to systematically
suppress all scattering events that would lead to signal velocities faster
than the speed of light, the second row of Fig.~\ref{fig9} summarizes the
results of this
simulation.

It is not surprising that the velocity restriction leads to a huge reduction in
the collision rate: While in the first stages of the collision the collision
rate drops to fourtynine percent of the original rate, it drops in the later
stages to only about eight percent of the original rate, there is no outwards
travelling information at all -- it would be unwise to apply such an
unjustified prescription given the huge effect, even more so, as from the
extrapolation of intermediate energy results one would expect a rather strong
outgoing effect of the proton impact.

\subsection{Suppressing of Low Energy Collisions}
Once a parton has scattered with a parton from the other nucleus it
can scatter with fellow partons from the same nucleus at much lower
c.m.-energies. Soft processes between partons from the same nucleus are
problematic since the elastic gluon-gluon cross sections at low energies are
large; systematically suppressing these reactions by placing a c.m.-energy
cut-off on the scattering events should also suppress high signal velocities.
For this model calculation the cross sections are set to zero below 0.4 GeV,
see the third row of Fig.~\ref{fig7}.
The outcome of this simulation proves that indeed the shock-wave travels
outwards through those low energy collisions, only the initial outburst
remains.

In fact this model provides a way to address the problem of the
energy-dependent parton resolution described in the beginning of
Sec.~\ref{sec4}: Disregarding interactions below a certain c.m.-energy cutoff
could also be viewed as disregarding the interaction partners. By smearing out
the step-function of the cutoff to a smoother function that determines the
probability of scattering events in dependence of the c.m.-energy, one could
try to simulate the behavior of the parton distribution function in dependence
of the resolution parameter $Q^2$.
However, other physics would have to take the
place of the QCD interactions, in this lower-energy regime nucleonic
interactions take over and hadronization patterns must be applied.

\subsection{First Steps Towards a Retarded Interaction}
As already pointed out, the main reason for the superluminous signal velocities
is the instantenous character of the elementary interactions.
To consequently overcome this problem no theoretical framework with time delays
has been developed so far for parton systems, although some simple hadronic
examples have been worked out. To estimate the effects of retardation it
nevertheless seems appropriate to introduce a simple -- but unphysical -- way
of time delay in the interactions: After an interaction took place the
momentum transfer to both partons involved is delayed by
$|\Delta\mbox{\boldmath$r$\unboldmath}_{\mbox{\scriptsize lab}}|/(2c)$,
where $\Delta\mbox{\boldmath$r$\unboldmath}_{\mbox{\scriptsize lab}}$ is the
distance vector in the lab
frame. In the meantime the partons are considered not being able to interact,
see Fig.~\ref{fig8}. A technical problem in our calculations however is that
the time delay has to be implemented for integer numbers of timestep lengths
and therefore does not necessarily fully correspond to its supposed length.

It turns out that the propagation of the outgoing shock-wave becomes more
homogeneous, the initial outburst is suppressed while the distances of the
outmost scattering events are comparable to the simulations with instantaneous
interactions, see the forth row of Fig.~\ref{fig7}.

\subsection{Downscaling the Cross Sections}
The maximum signal velocity is proportional to the maximum cross section.
Downscaling the cross sections to $1/n$th of its original value and
compensating this by the initialization of $n$ times as many testpartons leads
to a reduction of the maximum signal velocity by $1/\sqrt n$.
Due to limitations
given by the increasing computation time, for this model calculation we have
chosen a factor of only $n=5$ which corresponds to a reduction of the maximum
signal velocity by a factor of $\approx0.45$.

It turns out that this technique radically changes the characteristics of the
simulation, the outgoing shock-wave is strongly suppressed, see the fifth row
of Fig.~\ref{fig7}. This outcome is surprising, since Ref.~\cite{welke} found
that for internuclear cascade codes the outcome of this ``full-ensemble''
method ($n$ times as many test particles) is comparable the usual
``parallel-ensemble'' method ($n$ parallel test runs). In the case of parton
cascades apparently this does not hold true anymore due to the reduction of
superluminous signals.

Before favoring a method however that leads to such strong changes in the
outcome of the simulation, further analysis is mandatory.
\subsection{Wee-Partons}
When initializing the parton configuration the Lorentz-contraction leads to
very thin pancakes of nuclear matter, the contracted thickness of a nucleon is
around $8\cdot10^{-3}$ fm, that of the gold nucleus $6\cdot10^{-2}$ fm. Thereby
the $z$-coordinate of the partons is fixed rather precisely. However, the
momentum of the low-$x$-components of the nucleonic wave-function is also
determined within the order of a few hundred MeV. This leads to a violation of
the uncertainty principle, which lead to the suggestion that in any frame of
reference the nucleonic wave-function should be smeared out to a pancake
thickness of at least approximately 1 fm, where the low-$x$-components are
situated further outside and only the valence-quarks are actually within the
highly contracted pancake \cite
{geiger1,geiger2,geiger3,geiger4,geiger5,geiger6,mclerran}. Applying this kind
of initial configuration naturally leads to a smaller parton density; in our
case approximately by a factor of 20. Fig.~\ref{fig9} shows the parton
configuration both at the beginning of our simulation and at 2 fm/$c$.

One's hope can be that the increased width and smaller parton density of the
nuclei compared to the fully Lorentz-contracted model helps to eliminate some
causality violating effects, see the bottom row of Fig.~\ref{fig7}:
The initial outburst of a shock-wave at impact
is eliminated because the nuclei enter each other rather gradually, an outgoing
shock-wave within the respective nuclei is stronger damped because of a
decrease in collision rate, and finally time-ordering problems on the scale of
the time the nuclei are passing each other are rarified because that time-scale
becomes much longer.

However, there are now other inconsistencies: Although this concept given by
the uncertainty relation is truly valid in any
frame of reference, within this model there is no way of implementing it in a
Lorentz-invariant way. In reality the components of the wave-functions
transform, in different frames different components of the wave-function are
considered a parton. In our model only the parton coordinates and momenta
transform, while the partons themselves, once generated, exist in any frame.
While for the initial collisions the c.m.-frames of individual parton
collisions nearly coincides with the lab-frame, in subsequent collisions the
c.m.-frames are closer to the rest-frames of the respective nuclei
-- in its rest-frame however, a nucleon, smeared out to 1 fm in the lab-frame,
has a longitudinal radius of 100 fm.

\subsection{Proper time approach}\label{sec6f}
As already pointed out, a fully covariant description of a particle collisions
would require a fully 4-dimensional configuration space. This is not possible
without giving up the equal-time character of the simulation.
Ref.~\cite{kodama} however proposes a causality
preserving scheme that while retaining the unique global time character of the
simulation minimizes the frame-dependence of the choice of collision partners
for the particles, but not the frame-dependent time-ordering of those
collisions. Each particle $i$ is considered to have its own clock showing its
proper time $\tau_i$,
\begin{equation}
d\tau_i=dt\sqrt{1-\beta_i^2(t)},\quad\tau_i(t)=\int_0^t
dt'\sqrt{1-\beta^2_i(t')}\ .
\end{equation}
Since the particles do not change their momenta between collisions, the above
integral can be reduced to a sum of products of the type $\beta_{i\alpha}\Delta
t_\alpha$.
With $\tau_{ic}(j)$ being the proper $i$-time of the collision of
particle $i$ with particle $j$, and $\tau_{i0}$ being the proper $i$-time of
the most recent collision of particle $i$, let
\begin{equation}
\delta\tau_i(j):=\tau_{ic}(j)-\tau_{i0}
\end{equation}
be the proper time distance between those two events, a Lorentz-invariant
quantity. The collision instant $\tau_{ic}(j)$ is defined individually within
the rest-frame of particle $i$ through the closest approach to particle $j$ --
in the other methods, the closest approach is defined either within one frame,
usually the lab-frame or the c.m.-frame of the nuclei, or within the respective
c.m.-frame of a particle {\it pair}. In our previous discussions, we had chosen
the latter mechanism.
To co-relate the individual closest approach tests, for
collisions only the particle pairs $(i,j)$ are considered for which both
\begin{mathletters}
\begin{equation}
\delta\tau_i(j)=\mbox{Min}\left\{\delta\tau_i(l)>0,\ l=1,\ldots , N;
l\neq i\right\}\ \phantom{,}
\end{equation}
and
\begin{equation}
\delta\tau_j(i)=\mbox{Min}\left\{\delta\tau_j(l)>0,\ l=1,\ldots , N;
l\neq j\right\}\ ,
\end{equation}
\end{mathletters}
$N$ being the total number of particles in the simulation. Through the
restriction $\delta\tau>0$ only collisions in the absolute future of each
particle are considered. What the two above
conditions mean is that for particle $i$ the very next possible collision (in
the sense of proper time) is with particle $j$ {\it and vice versa}.
This algorithm only allows collisions that have no risk of not happening in
another frame. Suppose particle 1 has a minimum proper time distance to
particle 2, but particle 2 has its minimum proper time distance to
particle 3, then no collision will take place at all.

The search for the very next collision partner of every particle in the
simulation is an un-avoidable $N^2$-problem. Since each of these tests involves
a change of coordinate system, this mechanism is computationally very intense.
Therefore, it was only possible to simulate the initial stages of one single
(p,Au)-collision. We observed that the number of collisions in
this phase dropped dramatically, in our testrun to about 10\% of the number
in the other simulations; due to computational limitations however
we are not able to claim significant statistics on this percentage. This
outcome is compatible with the conjecture of the authors of Ref.~\cite{kodama}
that the mechanism might underestimate the number of collisions
\cite{kodamapriv}. In smaller simulations of
(p,p)-collisions situations were
observed where particle 1 had particle 2 as its closest collision partner,
particle 2 had particle 3, and particle 3 again had particle 1.
``Ring''-configurations like this, possibly spanning over even more particles,
might lead to the underestimation of the collision rate. This
effect might have been enhanced by the high particle density and the extreme
Lorentz-contraction of the nucleons as compared to lower energy internuclear
cascades the mechanism was developed for. Further investigation
of this method is required, but due to computational limitations was not
possible within this work. It is however not expected that the above mechanism
can overcome the macroscopic problem of
shock-waves.
\section{Conclusions}\label{sec7}
The influence of superluminous signal velocities on the signal propagation in
parton cascade codes was found to be smaller than expected. This is mainly due
to two effects: In the initial stages of the interaction the energy-dependent
cross sections tend to be small. Without this beneficial effect the signal
propagation has a threshold in the region between 0.4 and 0.5 fm$^2$ from where
on the velocity of the outgoing shock-wave reaches the speed of light.
In the later stages of the interaction shock-waves get damped out not because
of a lack of interactions but because of
signal propagation that resembles a random walk: not every collision actually
leads to an outward propagation, in fact, the equally probable inward
propagation leads to a virtual damping of the shock-wave.

In the near future the influence of particle production and particle absorption
will be examined.

\acknowledgments
We acknowledge useful discussions with C.-P. Yuan, W.-K. Tung, and
P. Danie\-lewicz. Also, we would like to thank T. Kodama and K. Geiger
for the helpful electronic mail exchange.
Research supported by an NSF presidential faculty fellow award and by
NSF grants 9017077 and 9403666, and by the Studienstiftung des Deutschen
Volkes (GK).

\begin{figure}
\caption[]{One particle coming from the left scatters with another particle
coming from the right. As part of the model the scattered particles cannot
scatter again in this timestep. However, one timestep later that scattered
particle from the
right scatters with another particle coming from the right.
As a result information from the first scattering, such as particle momenta
and type, has travelled to the second one. This kind of information
transport can continue over several scattering events.
\label{fig1}}
\end{figure}

\begin{figure}
\caption[]{Distance $d$ of scattering events versus simulation time at constant
cross sections of $\sigma=$0.1, 0.2 and 0.5 fm$^2$. The solid line
indicates the
distance that could be reached by a signal that travels with the speed of
light. One can clearly see the outwards travelling shock-wave and how it gets
damped out with time. For the smaller cross sections the causality violation
is rather small and only present in the initial stages of the interaction.
One should note that for realistic energy-dependent cross sections the initial
scattering events take place with a lower cross section than the later ones
because the c.m.-energy is much higher. In the lower right panel the maximum
distance $d_{\mbox{\scriptsize max}}$ of
scattering events from the beam axis is plotted versus the cross
section $\sigma$.
The horizontal line indicates a distance that during the simulation
time could only be reached with the speed of light.
\label{fig2}}
\end{figure}

\begin{figure}
\caption[]{Two partons passing their point of closest approach in the
c.m.-frame.
\label{fig3}}
\end{figure}

\begin{figure}
\caption[]{
Since a spacelike distance usually prevents causal dependencies of two
spacetime points, in the c.m.-frame the scattering of parton $A$ can happen so
much earlier than the scattering of parton $B$ that at the beginning of the
c.m.-frame's timestep (denoted by thin vertical lines in the figure)
parton $A$ has already scattered while parton $B$ has not
yet scattered at the end of that timestep. In the lab-frame however at the
beginning of its timestep parton $A$ and $B$ have not yet scattered while at
the end both have.
\label{fig4}}
\end{figure}

\begin{figure}
\caption[]
{Example of partial cross sections being used versus the c.m.-energy
$\sqrt s$. The gluon-gluon cross section is divergent for
$\sqrt s\to0$, a problem that cannot be solved with the introduction of
cut-off masses. The cross section  was chosen to be constant below
$s=0.25$ GeV$^2$ at a value of approximately $0.45$ fm$^2$.
\label{fig5}}
\end{figure}

\begin{figure}
\caption[]{In the top panel the projection of the initial parton configuration
on the $x,z$-plane is depicted. The highly Lorentz-contracted proton is moving
to the left into the equally contracted gold nucleus which is moving to the
right. The bottom panel shows the configuration 1 fm/$c$ later.
\label{fig6}}
\end{figure}

\begin{figure}
\caption[]{Results of simulations with different mechanisms to suppress
superluminous signal transport. The leftmost histogram shows the signal
velocity
distribution regarding directly subsequent collisions, $v_S$,
see Fig.~\ref{fig1}.
The solid histogram shows the radial component of those velocities, the dashed
histogram includes the longitudinal component.
The second panel shows a more general impression of the signal velocities, in
this histogram the signal velocity $v_A$ from the
very first to the very last scattering event that a parton was involved in
is calculated. Again, the solid histogram shows the radial component of the
signal velocities, the dashed histogram also takes into account the
longitudinal components. The third panel shows how
subsequent scattering events are leading to this overall damping of peak
velocities occurring in single scattering events: again starting from the
very first scattering event of a parton, the average signal velocity
$\langle v\rangle$ to the
$i$th following scattering event of the partons is calculated; both
the total and only the radial component of the velocities are shown.
Finally, the rightmost panel shows the outgoing shock-wave, in a plot of
distance from the beam axis versus simulation time the contour lines of the
scattering event distribution are given. For model A, the rejection of any
individual scattering events that would lead to superluminous signal transport,
the shock-wave was so strongly damped that it did not even go beyond the radius
of the incoming proton.
\label{fig7}}
\end{figure}

\begin{figure}
\caption[]{A first step to the introduction of retarded interactions: After the
scattering took place, the partons maintain their momentum for a time period
of $|\Delta$\boldmath$r$\unboldmath$|/(2c)$ in which they are not able to
scatter with other partons. Only after this time the momenta are transferred
to the partons.
\label{fig8}}
\end{figure}

\begin{figure}
\caption[]{Initial configuration and configuration after 2 fm/$c$
for a simulation with wee-partons.
\label{fig9}}
\end{figure}

\begin{table}
\narrowtext
\caption[]{Parameters used for the comparisons}
\begin{tabular}{l|l}
Parameter&Chosen Value\\
\tableline
Total time of simulation&3 fm/c\\
Timestep length&0.0002 fm/c\\
Number of timesteps&15000\\
Proton energy&100 GeV\\
Energy per nucleon Au&100 GeV\\
Impact parameter&0 fm\\
Minimum $x$-value for $f(x,Q^2)$&0.005\\
$Q$ for $f(x,Q^2)$&25 GeV\\
Bag-radius for nucleons&0.9 fm\\
$\alpha_s$&0.2\\
Cut-off mass for gluon propagators&1.0 GeV\\
Cut-off mass for quark propagators&0.2 GeV\\
Number of parallel test runs&5
\label{tab1}
\end{tabular}
\end{table}

\end{document}